\documentclass{PoS}

\usepackage{amsmath,amssymb}
\usepackage{cite}
\usepackage{xspace}
\usepackage{tikz}
\DeclareMathAlphabet{\mathcal}{OMS}{cmsy}{m}{n} 

\newcommand{\eqn}[1]{Eq.\,(\ref{#1})}

\newcommand{\fig}[1]{Fig.\,\ref{#1}}

\newcommand{\sct}[1]{Section~\ref{#1}}

\newcommand{\refes}[1]{Refs.\,\cite{#1}}

\title{Calculation of Multi-Loop Integrals with SecDec-3.0}

\ShortTitle{Calculation of Multi-Loop Integrals with SecDec-3.0}

\author{\speaker{Johannes Schlenk}, Tom Zirke\\
Max Planck Institute for Physics, F\"ohringer Ring 6, 80805 Munich, Germany\\
        E-mail: \email{jschlenk@mpp.mpg.de}, \email{zirke@mpp.mpg.de}}

\abstract{In this contribution we discuss new features of SecDec-3.0, a public program
for the evaluation of dimensionally-regulated parametric integrals using sector decomposition.
We will focus on two main aspects: the implementation of an improved geometric decomposition algorithm
and the recently added support for complex masses.
}

\FullConference{12th International Symposium on Radiative Corrections (Radcor 2015) and LoopFest XIV (Radiative Corrections for the LHC and Future Colliders)\\
		 15-19 June  2015\\
		 UCLA Department of Physics \& Astronomy Los Angeles, CA, USA}

\newcommand{\secdec}{\textsc{SecDec}\xspace}

\newcommand{\eps}{\epsilon}

\DeclareMathOperator{\real}{Re}
\DeclareMathOperator{\imag}{Im}
\newcommand*{\eg}{e.g.\@\xspace}
\newcommand*{\ie}{i.e.\@\xspace}

\newcommand{\tev}{~\text{TeV}\xspace}

\definecolor{grey}{RGB}{171,171,171}

\begin{document}

\section{Introduction}
As Run~II of the LHC at 13\tev is progressing and statistical uncertainties are decreasing,
precise predictions at higher loop orders involving many scales are needed to push the precision frontier.
An important building block of higher order predictions is the calculation of master integrals, which, however,
often cannot be performed analytically with current technology.

An alternative is offered by numerical methods, 
such as sector decomposition~\cite{Hepp:1966eg,Roth:1996pd,Binoth:2000ps,Heinrich:2008si},
which handles the $\eps$-expansion of dimensionally regulated multi-loop Feynman integrals and their numerical calculation.
The public program \secdec~\cite{Carter:2010hi,Borowka:2012yc,Borowka:2015mxa}, 
whose latest features we shall discuss in this paper, provides an automated implementation of this algorithm.
Other publicly available codes are sector\_decomposition \cite{Bogner:2007cr} and Fiesta~\cite{Smirnov:2008py,Smirnov:2009pb,Smirnov:2013eza,Smirnov:2015mct}.

The main part of this paper is structured as follows:
We begin with a recap of the sector decomposition algorithm in \sct{sec:secdec}.
\sct{sec:features} is dedicated to the new features of \secdec~3.0, 
with emphasis on the geometric decomposition strategy and the implementation of complex masses.
Finally we give our conclusion in \sct{sec:conclude}.

\section{Sector Decomposition}\label{sec:secdec}
Although the sector decomposition algorithm is applicable to a more general class of parametric integrals,
let us consider as a starting point of our discussion the Feynman parametrization of an $L$-loop integral,
 \begin{equation}I = \frac{(-1)^{N_{\nu}}}{\prod_{j=1}^{N}\Gamma(\nu_j)}\Gamma(N_{\nu}-LD/2)
 \int\limits_{0}^{\infty} 
\,\prod\limits_{j=1}^{N}\frac{\text{d}x_j}{x_j}\,\,x_j^{\nu_j}\,\delta(1-\sum_{l=1}^N x_l)\,\mathcal{N}\,
\frac{\mathcal{U}^{N_{\nu}-(L+1) D/2}}
{{\mathcal F}^{N_\nu-L D/2}},
\label{eq:feynint}
\end{equation}
where $\mathcal{U}$ and $\mathcal{F}$ are the graph polynomials, $\mathcal{N}$ is a numerator appearing in tensor integrals, and $D=4-2\eps$. The parameters $\nu_j$ denote the propagator exponents and $N_\nu=\sum_{j=1}^N\nu_j$.

The basic idea of the algorithm is to split up the integration region into multiple parts 
with non-overlapping singularities and perform a change of variables to map to the unit hypercube afterwards.
There exist several different decomposition algorithms, which will be discussed in more detail 
in \sct{sec:geomethod}.
As a result the singularities are isolated as factors of the form $x^{-a+b\eps}$ with $a>0$ so that they can be 
subtracted systematically. This makes it possible to perform an expansion in $\eps$.
Finally, the coefficients to each order in $\eps$ can be calculated numerically.

\section{New Features of SecDec-3.0}\label{sec:features}
The latest version of the program \secdec can be obtained from hepforge on the url:
\begin{center}
\texttt{http://secdec.hepforge.org/}
\end{center}
It requires a working version of Perl and Mathematica version 7 or higher, as well as a C++ compiler.

Users familiar with earlier versions of the program will notice the improved user interface.
It was designed with mainly two goals in mind: firstly to facilitate interfacing with reduction programs, \eg
by implementing support of zero and negative propagator exponents $\nu_j$, 
and secondly to simplify scans over ranges of kinematical parameters, which can now be done by simply 
providing a list of phase space points in the \texttt{kinem.input} file.

On the technical side, the list of available numerical integrators, so far consisting of Cuba~\cite{Hahn:2004fe,Hahn:2014fua} and Bases~\cite{Kawabata:1995th}, has been extended by Cquad~\cite{cquad} and the NIntegrate routine from Mathematica. Furthermore the compilation and numerical integration can now be parallelized in a straight-forward way on a cluster, where the submission systems condor and PBS are supported.

Also the field of possible applications was enlarged: Support for integrals with linear propagators following the Feynman $+i \delta$ prescription has been implemented, as well as support for complex masses. The latter will be discussed in \sct{sec:complexmass}. 
In addition, general parametric integrals may now contain $\eps$-dependent dummy functions.

Another major new feature, which shall be presented in the following, is the implementation of a geometric decomposition algorithm.

\subsection{Geometric Decomposition Algorithm}\label{sec:geomethod}
The heuristic decomposition algorithm implemented in \secdec suffers from the problem that in some cases an infinite recursion can occur and no full decomposition is obtained.
The algorithms described in \refes{Bogner:2007cr} and \cite{Smirnov:2008py} are guaranteed to terminate, but generically produce a larger number of sectors.
A different approach based on convex geometry was introduced by Kaneko and Ueda in \refes{Kaneko:2009qx,Kaneko:2010kj}. This algorithm is guaranteed to terminate, while also keeping the number of sectors small.

In addition to the heuristic algorithm, the most recent version of \secdec implements the original algorithm of Kaneko and Ueda ({\tt G1}) and an improved geometric decomposition algorithm ({\tt G2}).
Here we describe the improved geometric decomposition algorithm {\tt G2}.

In contrast to other sector decomposition algorithms where a primary sector decomposition is performed, the Cheng-Wu theorem~\cite{Cheng:1987ga,Smirnov:2006ry} is used to integrate out the Dirac delta in \eqn{eq:feynint}.
This amounts to replacing the $\delta$-distribution by $\delta(1-x_N)$.

In the next step the Newton polytope $\Delta$ of the polynomial $\mathcal{U}\cdot\mathcal{F}\cdot\mathcal{N}=\sum_{j=1}^m c_j\mathbf{x}^{\mathbf{v}_j}$ is calculated, which is defined as the convex hull of the $(N-1)$-dimensional exponent vectors $\mathbf{v}_j$:
\begin{equation}
\Delta=\text{ConvHull}(\mathbf{v}_1, \dots ,\mathbf{v}_m).
\label{eq:vertex}
\end{equation}
Here the multi-index notation $\mathbf{x}^{\mathbf{v}_j}=\prod_ix_i^{(\mathbf{v}_j)_i}$ is used. The polytope $\Delta$ contains all necessary information for the sector decomposition. Due to the Minkowski-Weyl theorem there exists a second representation of the polytope $\Delta$ as an intersection of halfspaces defined by the facet normal vectors $\mathbf{n}_F$ \cite{Oda1988}:
\begin{equation}
\Delta = \bigcap_{F} \left\{ \mathbf{x}\in\mathbb{R}^{N-1} \mid \langle \mathbf{x},\mathbf{n}_F\rangle + a_F \geq 0 
\right\}\;.
\label{eq:facet}
\end{equation}
In \secdec the program {\sc Normaliz 2.10.1}~\cite{2012arXiv1206.1916B,Normaliz} is used to translate between vertex and facet representations, \eqn{eq:vertex} and \eqn{eq:facet}, 
respectively.

For each extremal vertex of $\Delta$ indexed by the parameter $j$, a sector is introduced. The sector is bounded by the facet vectors $\mathbf{n}_F$ incident to the vertex $j$. In order to map the integration region back to the unit hypercube, the local change of variables 
\begin{equation}
x_i=\prod_{F\in S_j} y_F^{\langle \mathbf{e}_i,\mathbf{n}_F \rangle}
\label{eq:trafo}
\end{equation}
is performed in sector $j$. The vectors $\mathbf{e}_i$ denote the orthonormal basis of $\mathbb{R}^{N-1}$, the set $S_j$ contains the facets incident to the vertex $j$.
In cases where the set $S_j$ contains more than $N-1$ elements, an additional triangulation of the sector is needed.
In \secdec the triangulation algorithm implemented in {\sc Normaliz} is used for this purpose.

Compared to the other strategies implemented in \secdec, strategy {\tt G2} is the fastest method and it usually produces the smallest number of sectors.\\

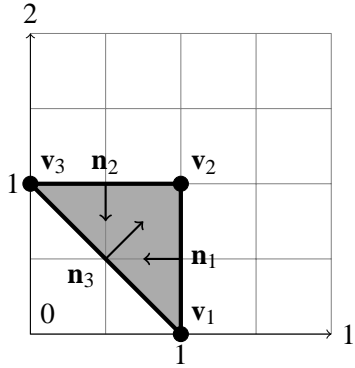
\begin{figure}
\centering
\begin{tikzpicture}
\draw[fill=grey] (2,0) -- (2,2) -- (0,2) -- (2,0);
\draw [help lines] (0,0) grid (4,4);
\draw[ultra thick] (2,0) -- (2,2) -- (0,2) -- (2,0);
\draw [<->] (4,0) -- (0,0) -- (0,4);
\node [right] at (4,0) {$1$};
\node [above] at (0,4) {$2$};
\node [left] at (0,2) {$1$};
\node[above right] at (0,0){$0$};
\node [below] at (2,0) {$1$};
\draw[fill=black] (2,0) circle [radius=0.1];
\draw[fill=black] (2,2) circle [radius=0.1];
\draw[fill=black] (0,2) circle [radius=0.1];
\node[above right] at (2,0){$\mathbf{v}_1$};
\node[above right] at (2,2){$\mathbf{v}_2$};
\node[above right] at (0,2){$\mathbf{v}_3$};
\draw[->,thick] (1,1) -- (1.5,1.5);
\draw[->,thick] (2,1) -- (1.5,1.0);
\draw[->,thick] (1,2) -- (1.0,1.5);
\node[below left] at (1,1){$\mathbf{n}_3$};
\node[right] at (2,1){$\mathbf{n}_1$};
\node[above] at (1,2){$\mathbf{n}_2$};
\end{tikzpicture}
\caption{Newton polytope $\Delta$ associated to the two loop vacuum integral of \protect\eqn{eq:vac}}
\label{fig:newton}
\end{figure}

As an example we decompose the two-loop vacuum integral with one massive and two massless propagators using strategy {\tt G2}.
After employing the Cheng-Wu theorem to integrate out the massive Feynman parameter $x_3$, the Feynman integral becomes
\begin{equation}
I=\begin{tikzpicture}[scale=0.75,baseline={([yshift=-.5ex]current bounding box.center)}]
   \draw [thick,domain=0:180] plot ({cos(\x)}, {sin(\x)});
   \draw [dashed,thick,domain=180:360] plot ({cos(\x)}, {sin(\x)});
   \draw [dashed,thick] (-1,0) -- (1,0);
\draw[fill=black] (-1,0) circle [radius=0.05];
\draw[fill=black] (1,0) circle [radius=0.05];
\node[above] at (0,1){$m$};
\end{tikzpicture} =-\Gamma(-1+2\eps)\left(m^2\right)^{1-2\eps} \int_0^{\infty} \frac{\text{d} x_1 \text{d} x_2}{ \left(x_1^1x_2^0+x_1^1x_2^1+x_1^0 x_2^1\right)^{2-\epsilon}}. 
\label{eq:vac}
\end{equation}
The exponent vectors
\begin{equation}
{\mathbf{v}_{1}}=
\begin{pmatrix}
1\\
0
\end{pmatrix}, 
{\mathbf{v}_{2}}=
\begin{pmatrix}
1\\
1
\end{pmatrix}, 
{\mathbf{v}_{3}}=
\begin{pmatrix}
0\\
1
\end{pmatrix}
\end{equation}
can be read off from the polynomial in the denominator of \eqn{eq:vac} and the associated Newton polytope $\Delta$ is shown in \fig{fig:newton}. 

The facet normal vectors 
\begin{equation}
\begin{matrix}
\mathbf{n}_1=&
\begin{pmatrix}
-1\\
0
\end{pmatrix} &
\mathbf{n}_2=&
\begin{pmatrix}
0\\
-1
\end{pmatrix} &
\mathbf{n}_3=&
\begin{pmatrix}
1\\
1
\end{pmatrix} \\
a_1=&1 &
a_2=&1 &
a_3=&-1 
\end{matrix}
\end{equation}
together with \eqn{eq:facet} specify the facet representation of the polytope $\Delta$. 
The sets $S_j$ associated to the three extremal vertices $\mathbf{v}_1$ to $\mathbf{v}_3$ are $S_1=\{3,1\}$, $S_2=\{1,2\}$ and $S_3=\{2,3\}$.
In this case no additional triangulation is necessary since the size of the sets already equals $N-1$.
The change of variables defined in \eqn{eq:trafo} can then be written as 
\begin{equation}
\begin{matrix}
x_1=y_1^{-1}y_3,\\
x_2=y_2^{-1}y_3
\end{matrix}
\end{equation}
leading to the decomposed form of the vacuum integral
\begin{equation}
I=-\Gamma(-1+2\eps)\left(m^2\right)^{1-2\eps}\int_0^1 \text{d} y_1 \text{d} y_2 \text{d} y_3\,\frac{y_1^{-\epsilon}\,y_2^{-\epsilon}\,y_3^{-1+\epsilon}}{\left(y_1+y_2+y_3 \right)^{2-\epsilon}} \left[ {\delta(1-y_2)} + {\delta(1-y_3)} + {\delta(1-y_1)} \right],
\end{equation}
where the $\delta$-distributions correspond to the sets $S_1$ to $S_3$.

\subsection{Complex Masses}\label{sec:complexmass}
In certain applications, especially in the electroweak context, the width of unstable particles can be important.
A consistent treatment is provided by the complex-mass scheme~\cite{Denner:1999gp,Denner:2005fg}, 
where the width $\Gamma$ is included as a negative imaginary part of the mass via the replacement
\begin{align}
m^2 \to m_c^2 \equiv m^2\left(1-i\frac{\Gamma}{m}\right).
\end{align}
The graph polynomial $\mathcal{F}$ then has the form
\begin{align}
\mathcal{F} = \mathcal{F}_0 + \mathcal{U} \sum_j x_j \left( m_j^2 - i  m_j \Gamma_j  \right),
\end{align}
\ie the widths induce a negative imaginary part:
\begin{align}
\imag\mathcal{F} = -\mathcal{U} \sum_j x_j  m_j \Gamma_j
\end{align}

In general, for zero widths, $\mathcal{F}$ will exhibit kinematic-dependent zeros even after sector decomposition, 
which can be avoided by a suitable deformation of the integration contour~\cite{Soper:1999xk,Nagy:2006xy,Binoth:2005ff}.
Similarly, a non-zero width can help to avoid these singular regions as well, but
one cannot expect this to lead to a stable numerical integration in all cases.
Thus it makes sense to try to combine the two in a consistent way, which should be possible since both the contour deformation and the complex masses are required to produce only negative imaginary parts in order to fulfill the Feynman $+i\delta$~prescription.
For \secdec-3.0 we have chosen
\begin{subequations}
\begin{align}
\vec{z}(\vec{x}) &= \vec{x} - i \vec\tau(\vec{x}), \\
\tau_k &= \lambda x_k (1-x_k)  \frac{\partial\mathcal{\real F}}{\partial x_k},
\end{align}
\end{subequations}
\ie to set the widths to zero in the definition of the deformation.
For small deformations we then have
\begin{align}
  \mathcal{F}(\vec{z}(\vec{x})) &= \real\mathcal{F}(\vec{x}) + i \imag\mathcal{F}(\vec{x})
-i\lambda \sum_k x_k(1-x_k) \left[\left(\frac{\partial\mathcal{\real F}}{\partial x_k}\right)^2
 + i \frac{\partial\real\mathcal{F}}{\partial x_k}\frac{\partial\imag\mathcal{F}}{\partial x_k}\right]
\nonumber\\
&\quad- \frac{\lambda^2}{2} \sum_{k,l} x_k(1-x_k) x_l (1-x_l)
\frac{\partial\mathcal{\real F}}{\partial x_k} \frac{\partial\mathcal{\real F}}{\partial x_l} 
 \left[ 
\frac{\partial^2\real\mathcal{F}}{\partial x_k \partial x_l} 
+i\frac{\partial^2\imag\mathcal{F}}{\partial x_k \partial x_l} 
\right]
+ \mathcal{O}(\lambda^3).
\end{align}
Up to order $\lambda$, the imaginary parts induced by the widths and the contour deformation are both negative
as they should. The term involving $\frac{\partial\imag\mathcal{F}}{\partial x_k}$ does no harm
because it is purely real.
At order $\lambda^2$, however, $\imag\mathcal{F}$ leads to an imaginary part of indefinite sign, which 
would otherwise have been the case at one order higher in $\lambda$.
On the other hand, this term is proportional to $\frac{\Gamma_j}{m_j}$ and thus suppressed 
since the widths should be small compared to the corresponding masses.
Therefore we conclude that for a sufficiently small value of $\lambda$, one can consistently combine complex masses and contour deformation.

The support for complex masses is included in \secdec from version~3.0.8
and can be enabled by setting \texttt{complexmasses=1} in the \texttt{param.input} file.
It may be used with and without contour deformation.
If complex masses are switched on, the \texttt{kinem.input} file expects two numbers for each mass parameter:
$$ \cdots \, \real m_{c,1}^2 \,\, \imag m_{c,1}^2 \,\, \real m_{c,2}^2 \,\, \imag m_{c,2}^2 \, \cdots $$
As an example the calculation of a one-loop pentagon with complex masses can be found in the \texttt{loop/demos/11\_complexmass} folder.

\section{Conclusions}\label{sec:conclude}
The new version 3.0 of the program \secdec comes with an improved user interface and opens up several new possibilities:
Two geometric decomposition strategies, which are guaranteed to stop, have been implemented, as well as support for linear propagators, complex masses, $\eps$-dependent dummy functions, and new integration routines.
Furthermore, the more flexible definition of integrals facilitates the link to reduction programs. 
There is an option to scan over phase-space points, and the new cluster mode makes it easier to perform calculations on modern super computers.

With these features, \secdec is now ready for the application to phenomenologically relevant processes and has the potential to become an important building block in a future automated two-loop setup.

\section*{Acknowledgments}
We would like to thank the members of the \secdec team, Sophia Borowka, Gudrun Heinrich, Stephan Jahn, Stephen Jones, and Matthias Kerner for the successful collaboration and helpful comments on the manuscript.

\bibliographystyle{JHEP}

\bibliography{refs_secdec}

\end{document}